\documentclass[12pt]{iopart}
\usepackage{graphicx}
\usepackage{cite} 
\usepackage{xcolor}
\usepackage{hyperref}
\usepackage{amssymb}
\usepackage{tikz}
\hypersetup{colorlinks=true,linkcolor=cyan,urlcolor=cyan,citecolor=cyan,}
\usepackage[pagewise]{lineno}

\begin{document}
\title[Local equilibrium properties of  the Sinai diffusion]{Local equilibrium properties 
	of ultraslow diffusion in the Sinai model}

\author{Amin Padash$^{1,*}$,  Erez Aghion$^2$,  Alexander Schulz$^1$, Eli Barkai$^3$, Aleksei V. Chechkin$^{4,5,6}$,  Ralf Metzler$^{4,7}$ and Holger Kantz$^1$}

\address{$^1$ Max Planck Institute for the Physics of Complex Systems,
N{\"o}thnitzer Stra{\ss}e 38, D-01187 Dresden, Germany}
\address{$^2$ Departments of Physics and Chemistry, University of Massachusetts Boston, MA 02125, USA}
\address{$^3$ Department of Physics, Bar Ilan University, Ramat-Gan 52900, Israel }
\address{$^4$ Institute of Physics \& Astronomy, University of Potsdam, D-14776
Potsdam-Golm, Germany}
\address{$^5$ Faculty of Pure and Applied Mathematics, Hugo Steinhaus Center,
Wroclaw University of Science and Technology, Wyspianskiego 27, 50-370
Wroclaw, Poland}
\address{$^6$Akhiezer Institute for Theoretical Physics, 61108 Kharkov, Ukraine}
\address{$^7$ Asia Pacific Centre for Theoretical Physics,  Pohang 37673, Republic of
Korea}
\address{$^*$ Author to whom any correspondence should be addressed.}
\ead{padash@pks.mpg.de}

\vspace{10pt}
\begin{indented}
\item[]
\end{indented}

\begin{abstract}
We perform numerical studies of a thermally driven, overdamped particle in a random quenched force field, 
known as the Sinai model. We compare the unbounded motion on an infinite 1-dimensional domain to the motion in bounded domains with reflecting boundaries and show that the unbounded motion is 
at every time close to the equilibrium state of a finite system of growing size. 
This is due to time scale separation:
Inside wells of the random potential, there is relatively fast equilibration, while the motion across major potential barriers is ultraslow. Quantities studied by us are the time dependent mean squared displacement, the time dependent 
mean energy of an ensemble of particles, and the time dependent entropy of the probability distribution. 
Using a very fast numerical algorithm, we can explore times up top $10^{17}$ steps and thereby also 
study finite-time crossover phenomena. 
\end{abstract}

\vspace{2pc}
\submitto{\NJP}
\maketitle

\section{Introduction}
Among many models for subdiffusion, the Sinai model sticks out due to 
the fact that diffusion is ultraslow. This means that the mean squared
displacement grows slower than any power of time, namely like $\ln^4 t$,
 see \cite{Liangetal} for a recent review on other such systems.
 Physically, the Sinai model describes the one-dimensional thermal random motion 
of a particle in a random
potential, more specifically, a potential which is constructed from a Brownian
path. If the spatial domain is infinite, as it is usually assumed when writing
down the model, the potential can have arbitrarily high barriers and
arbitrarily deep wells, however, due to the recurrence properties of Brownian
paths in one dimension, there will be potential zero-crossings at arbitrarily far
distances from the origin. 

This model was introduced by Yakov Sinai \cite{Sinai1982} 
as  a special case of models with a site-dependent jump probability and has found much attention 
in the literature since then, see, e.g., \cite{Kesten1986,Nauenberg1985, Bunde1988, Comtet1998, bouchaudetal,Bouchaud1990} 
for some thorough analysis. The concept has found many applications, including the dynamics of random field magnets and dislocation dynamics \cite{bouchaudetal}, glass dynamics \cite{Doussal1995}, aging phenomena \cite{Laloux1998}, random-field Ising models \cite{Fisher2001, Bruinsma1984} and helix-coil boundaries in random heteropolymers \cite{Gennes1975, Oshanin2009}. With the inherently quenched heterogeneity of biomolecules, Sinai-type models describe mechanical DNA unzipping \cite{Walter2012, Kafri2006}, translocation of biopolymers
through nanopores \cite{Mathe2004, Lubensky1999}, and molecular motors \cite{Kafri2004}. Also, quantum transport in disordered topological quantum wires \cite{Bagrets2016} has been related to the Sinai model. More generally, there are many phenomena of ultraslow diffusion in disordered systems of low dimension, such as in vacancy-induced motion \cite{Brummelhuis1988, Benichou2002}, biased motion in exclusion processes \cite{Juhasz2005}, local relaxation dynamics in DNA \cite{Brauns2002}, paper crumpling under a heavy piston \cite{Matan2002} or compaction of granular systems \cite{Richard2005}, glassy systems
\cite{Boettcher2011}, statistics of extreme events \cite{Schmittmann1999}, the ABC model \cite{Afzal2013}, dynamics in nonlinear maps \cite{ Drager2000},  interacting many-particle systems \cite{Sanders2014}, in dynamics of cooling granular gases \cite{Brilliantov, Bodrova2015-1}, and short range correlated Gaussian potentials \cite{Goychuk}. 

Analytical models describing such type of
motion include continuous time random walks with ultraheavy tailed distributions of
waiting times \cite{Drager2000, Godec2014, Chechkin2017}, ultraslow scaled Brownian motion \cite{Bodrova2015-2, Bodrova2016}, ageing continuous time
random walks \cite{Lomholt2013}, diffusion processes with strongly localised diffusivity \cite{Cherstvy2013, Cherstvy2013-2}, and distributed order fractional
diffusion equations \cite{Chechkin2003, Sandev2015, Sandev2016}, see also \cite{Metzler2014} for more references. 
Moreover, generalisations of the Sinai model including the presence of a fixed bias \cite{bouchaudetal},
random local bias \cite{Fisher1998,Fisher2001,Selinger1989,Woods2010},  
correlated and periodic disorder
\cite{Dean2014, Oshanin2013, Dean2016} have been studied.
The first passage time, persistence probability and the splitting probability
of the Sinai model is also reported in \cite{Comtet1998,Majumdar2002,Oshanin2009}.

Previous works concern the properties of a random walk in the Sinai model, for example the probability density function (PDF), transport phenomena and the mean squared displacement, based on long-time disorder-averaged dynamics by different approaches such as scaling arguments \cite{bouchaudetal, Goychuk}, the renormalisation group technique  \cite{Fisher1998} and a discrete random walk model \cite{Oshanin2013,Chave1999,Radons2004} as well as approaches based on time-averaged observables \cite{Godec2014}. However,
experimental observations 
often relate to non-equilibrium properties of the system, which is in the focus of the present paper.
More specifically, based on a discrete random walk model, we look into correlation phenomena and ageing when starting
ensembles of independent 
trajectories in the same realisation of the random potential 
but with their individual thermal 
noises by analysis of the ensemble mean potential energy and of the Shannon entropy of
the time dependent probability density. We then average over the disorder of
the potential. In our setup all particles are starting from the origin and also the potentials performing a random walk from this point to both directions. The main conclusion of this work is that due to the
extreme slowness of the diffusion, the system is always in quasi-equilibrium
on the domain explored so far. This is formalised by the concept of infinite,
non-normalised densities which was developed for non-confining, i.e.,
asymptotically one dimensional flat potentials \cite{Aghion2019}, also for log potentials \cite{Aghion2020}, and which
allows one to represent a time
dependent density by the invariant density on a finite domain dressed by a
time dependent factor.

For Sinai diffusion a remarkable result was obtained by Golosov: for a given realisation of a potential landscape all trajectories with the same initial
condition in the deepest well stay close together forever \cite{Golosov, Monthus2002, Monthus2003, Doussal2003}.
More practically, Golosov proved that in the long time limit, the distribution of the relative distance $y=x(t)-m(t)$, where $m(t)$ is the most probable position, after averaging over random potentials, tends to a limit distribution \cite{Golosov}.
However, by the renormalisation group analysis it was shown that for the same thermal initial conditions the existence of the limit distribution for the random variable $y$, does not imply that its moments remain
finite in the long time limit \cite{Monthus2002, Monthus2003, Doussal2003}. Here we scrutinise the case of non-equilibrium
initial conditions and show that quasi-equilibrium states in the random
potential emerge due to a time scale separation with respect to escape
times over larger local potential maxima. In contrast to the Golosov
result, however, these "local equilibria" depend on the starting point
of the trajectory and are non-universal in this sense.

Our paper is organised as follows. In section 2 we introduce an ergodic lattice hopping model to study the dynamics of the Sinai diffusion, followed by a numerical scheme using Markov matrix approach in section 3. Sections 4 and 5 reports the time dependent properties of the Sinai model such as the mean squared displacement, mean potential energy and the Shannon entropy, respectively. Equivalence of an unbounded motion in an infinite system and the equilibrium sate of a finite system is then investigated in
section 6. We discuss the different energy regimes of the infinite system with open boundary condition (open system) and the finite system with reflecting boundary condition (closed system) in section 7, and a comparison with infinite densities presented in section 8. Finally, a summary and discussion is provided in the last section.

\section{Lattice Sinai model}

\begin{figure}
\centering
\includegraphics[width=0.7\textwidth]{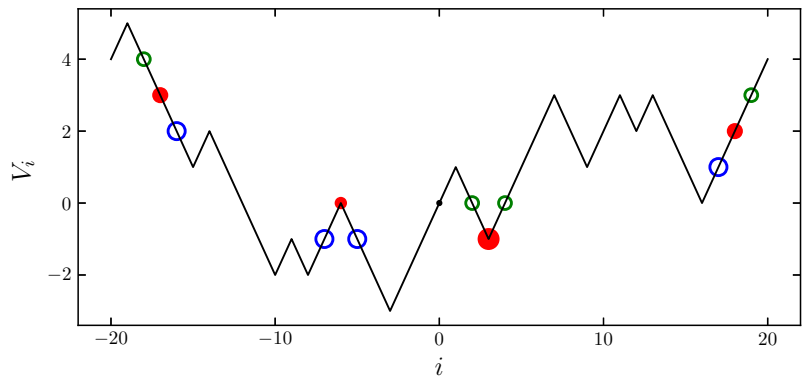}
\caption{\label{figure 1} Schematic of an arbitrary potential landscape and the hopping rate probabilities at different points of the potential. 
}
\end{figure}

We study a version of the Sinai model on the infinite lattice of integers, also called lattice hopping model \cite{Alexander1981, Hughes_book}. 
To create the random potential, we start a lattice random walk at site $i=0$ 
into both directions,
i.e., we define a path $V_i$ with $V_0=0$ and $|V_{i\pm1}-V_i| = 1$, so that
from 
one lattice site to its neighbour, $V_i$ jumps upward or downward by one unit
with equal probability. Then  
$V_i$ forms a lattice Brownian path in $i$, which here is a spatial coordinate.
We then interpret $V_i$ to be the potential for an
overdamped particle which is driven by thermal noise.  
The balance between the
deterministic downhill motion of an overdamped particle in this potential $V$
and thermal noise is determined by the parameter $\epsilon\in[0,1]$, 
which enters in the
probability for the particle at site $i$ to hop either to the left or to the
right or to stay:

\begin{eqnarray}\label{eq:jumpprob}
	q_{i\to i+1} &=& \frac{1}{4} + \frac{\epsilon}{4} (V_{i}-V_{i+1})\;,\nonumber\\
	q_{i\to i-1} &=&  \frac{1}{4} + \frac{\epsilon}{4} (V_{i}-V_{i-1})\;,\nonumber\\ 
	q_{i\to i} &=& 1 - q_{i\to i-1}- q_{i\to i+1} = \frac{1}{2}
	+ \frac{\epsilon}{4}(V_{i+1}-2V_i+V_{i-1})\;, 
\end{eqnarray}
where evidently the sum $q_{i\to i+1} + q_{i\to i-1} +q_{i\to i} =1$ 
irrespective of $\epsilon$. Since $|V_i-V_{i\pm1}|=1$, the probabilities to jump 
to a neighbouring site is given by one of the two values $1/4-\epsilon/4$ (uphill, \tikz\draw[green] (0,0) circle (.45ex);)
and $1/4+\epsilon/4$ (downhill, \tikz\draw[blue] (0,0) circle (.7ex);), 
while the probability to stay is one of the three values $1/2$ (\tikz\draw[red,fill=red] (0,0) circle (0.5ex);),
$1/2-\epsilon/2$ (\tikz\draw[red,fill=red] (0,0) circle (.35ex);), $1/2+\epsilon/2$ (\tikz\draw[red,fill=red] (0,0) circle (.8ex);), if the site $i$ is either on a slope, at a
maximum, or at a minimum of the potential, respectively (see figure~\ref{figure 1}). This restricts $0\le
\epsilon\le 1$, where $\epsilon = 0$ is the infinite temperature limit,
in which the potential does not influence the hopping rates, and $\epsilon=1$ is
the zero-temperature limit in which only downhill motion and resting, namely deterministic motion, with no fluctuations are possible 
and no particle can escape a potential minimum.

In contrast to implementations of this model without resting probabilities,
i.e.\ $q_{i\to i}=0$,  
this version is ergodic (no splitting into even and odd sub-lattices)\cite{ Hughes_book,Oshanin2013}. 
The even bigger advantage is that, as we will show below, equation~(\ref{eq:jumpprob})
has an exact analytical solution if we restrict the dynamics to a finite
domain with reflecting boundary conditions. Before we do so, let us 
discuss a special, non-random potential with infinite walls which has 
a V-shape: $V_i=|i|$. The stationary probability distribution $\mathrm{P}_i$ is
given through the detailed balance condition as $\mathrm{P}_i q_{i\to i+1} = \mathrm{P}_{i+1}
q_{i+1\to i}$. We compare this to the analytical solution of
the continuous in time and space Fokker-Planck equation with the potential $V(x)=c|x|$, where the constant $c$ carries the physical units. It 
is the Boltzmann distribution $\mathrm{P}^{B}(x)=Z^{-1} \mathrm{exp}\left({-c|x|/k_{B}T}\right)$, where $Z$ is the normalising partition function, $k_B$ is the Boltzmann factor, and 
$T$ is the temperature. Note that the diffusion coefficient reads $D=k_{B}T/\eta$ where $\eta$ is the damping, due to the Einstein relation. For a lattice point $i$ in the
slopes of this potential, the ratio of the hopping rates is the ratio $(1-\epsilon)/(1+\epsilon)$ of
uphill probability and downhill probability. The same ratio evaluated for 
the stationary solution of the Fokker-Planck equation is simply $\mathrm{e}^{-c|a|/k_{B}T}$, with the lattice size $a=1$, which together yields 
\begin{equation}\label{eq:temp}
	\frac{1-\epsilon}{1+\epsilon} = \mathrm{e}^{-c/{k_{B}T}} \mbox { or } k_{B}T/c = \ln^{-1}\left(\frac{1+\epsilon}{1-\epsilon}\right).
\end{equation}
Hence, we can relate $\epsilon$ also
quantitatively to temperature and as stated before, $T=0$ corresponds to
$\epsilon=1$ and $T=\infty$ to $\epsilon=0$. In the numerical simulations we set $c=1$. 

If we consider the system with its random potential on a finite domain $i\in[-L,L]$ with reflecting
boundary conditions, then an analytical expression for its invariant density $\mathrm{P}_i$ 
can be shown to be the Boltzmann distribution. The derivation works as follows:
The invariant distribution $\mathrm{P}_i$ satisfies detailed balance, 
$\mathrm{P}_i q_{i\to i+1} = \mathrm{P}_{i+1} q_{i+1\to i}$.  We fix arbitrarily the value ${\mathrm{P}}_0=1$ and
normalise all ${\mathrm{P}}_i$ after we have calculated them. The precise
discrete space solution can be easily obtained in the following way: 
$\mathrm{P}_{i+1}=\mathrm{P}_i q_{i\to
	i+1}/q_{i+1\to i}$ and hence $\mathrm{P}_{i+1} = \mathrm{P}_0 \prod_{k=0}^i q_{k\to
	k+1}/q_{k+1\to k}$. Inserting the transition
probabilities defined in equation~(\ref{eq:jumpprob}) we see that 
every ratio $q_{k\to k+1}/q_{k+1\to k}$ can only assume 
the value $(1-\epsilon)/(1+\epsilon)$ or its inverse,
depending on whether the jump is uphill or
downhill. Hence, in the
product, an equal number of uphill and downhill jumps cancel out their contributions, so that 
the result is:
\begin{equation}\label{eq:boltzmann}
	\mathrm{P}_i = \mathrm{P}_0 \left(\frac{1-\epsilon}{1+\epsilon}\right) ^ {V_i-V_0} = \mathrm{P}_0 \,\mathrm{e}^{-(V_i-V_0)/k_{B}T},
\end{equation}
where the latter is a consequence of equation~(\ref{eq:temp}). So we see that the
equilibrium probabilities follow a Boltzmann distribution which can be
normalised for finite $L$ by adjusting $\mathrm{P}_0$. These $\mathrm{P}_i$s  
can be calculated numerically with high accuracy so that we can numerically
evaluate all kinds of averages in thermodynamic equilibrium. The above
calculation represents a closed system with reflecting boundaries, since there is no in- or outflow of probability to lattice sites outside $[-L,L]$.

In order to study the dynamics of this model, 
in a straightforward numerical simulation one would first generate the 
random potential $V_i$ and then iterate a trajectory 
by random jumps from one lattice site to one of its neighbours according to
the probabilities of equation~(\ref{eq:jumpprob}). Repeating this many times for the same
random potential, one would simulate an ensemble of non-interacting particles
from which one can approximate time dependent position distributions. We are
interested in such distributions as a function of time. However, since the
diffusion in this potential is extremely slow, we need a much faster iteration
scheme, so that we are able to average also over many random potentials.

\section{Numerical scheme for the non-equilibrium system}
In the following, we focus on the initial condition $\mathrm{P}_i(t=0) =
\delta_{i,0}$, where all particles start at the lattice site $i=0$. 
Instead of time
consuming single particle simulations, we use a Markov matrix approach:
The discreteness of our physical space and time allows us to summarise 
a single step in the 
time evolution of the probability $\mathrm{P}(t)$ by the
multiplication of a Markov matrix with the vector $\mathrm{P}(t)$, where 
the elements of the Markov matrix are the transition probabilities $q_{i\to j}$ from one
lattice site to any other. These transition probabilities are given by
equation~(\ref{eq:jumpprob}) and hence the Markov matrix ${\cal M}$ has nonzero
entries only on the diagonal and the two secondary diagonals. 
Hence, $\mathrm{P}(t+1)={\cal
	M}\mathrm{P}(t)$. Instead of performing this multiplication one-by-one in
time, we simply take squares of the actual Markov matrix and thereby create
a sequence of matrices ${\cal M}^2$, ${\cal M}^4$, ${\cal M}^8$, etc. which generate $\mathrm{P}(t_k)$, where $t_k=2^k$, by
$k$ matrix multiplications. Even though a single such operation scales like
$N^3$ where $N$ is the rank of the matrix, we can quickly achieve large
$t_k$. If the initial condition is $\mathrm{P}_0(t=0)=1$ and $\mathrm{P}_{i\ne 0}(t=0)=0$, 
then the distribution at time $t_k=2^k$ is simply the central row of ${\cal
	M}^{2^k}$, $\mathrm{P}_i(t=2^k)= ({\cal M}^{2^k})_{0i}$. Since there is a non-zero
probability that a particle hops one step to the right in every iteration
step, after $2^k$ time
steps in principle a range of $2^k$ lattice points might be explored by a
trajectory. However, this probability, although theoretically strictly
non-zero, in practice is extremely small. 

While the disorder-averaged mean squared displacement MSD grows very slowly in
time, the motion in a single realisation of the potential is more complicated. Actually, as we
will illustrate later in more detail, particles and also the time dependent
probability distribution explore the lattice in a highly intermittent way.
If a particle is trapped (the probability is localised) in a deep potential
well, then for a long time the lattice will not be explored any further, only
on a much larger time scale we will see hopping to an even deeper well farther
away from the origin. This slowness implies
that on average over many such potentials the exploration horizon grows only like $\ln^2 t$ \cite{bouchaudetal,Bouchaud1990}. This is also true for most individual realisations of 
the random potential, so that 
the rank of the matrix $2L+1$, which is given by the range $-L:L$ of the lattice which we
model explicitly, can be chosen much smaller than $N= 2^{k_{max}}$. 
Since our (truncated) Markov
Matrix ${\cal M}$ is {\sl not} conserving probability (there is leakage 
out of the finite range of the lattice which we consider), eventually, the
norm of $\mathrm{P}$ will decrease. We can detect this numerically, and we will stop the time evolution when this leakage exceeds a total probability of 0.01. After which time this occurs depends on the 
individual realisation of the random potential, on the size of the resolved domain, and on the temperature $T$ or the noise strength $\epsilon$, respectively. In all the analyses below, we therefore vary the modelling range $[-L,L]$ 
of the open system, and we show results only up to times for which the leakage of probability was sufficiently small. For comparison, we also include results obtained for closed systems (reflecting boundaries) on equally large domains, obtained by the same numerical scheme.

On a standard workstation we thereby
arrive at $t > 10^{17}$ time steps for an ensemble of $10^4$ potential
landscapes. Actually, there is another problem besides leakage, which prevents us
from going to much larger times: numerical inaccuracy. For $k$ being larger
than 58, the matrix elements are very heterogeneous in their magnitude, so that
under squaring such a matrix, we add very large and very small numbers. 
In such a summation, the small numbers tend to be truncated by round-off. This
expresses itself in resultant matrices which from some number of 
squaring onward violate the normalisation of probability considerably,
independent of leakage. Hence, we trust our simulations only up to
$2^{58}\approx 10^{17}$ time steps. 

Direct numerical validations for this algorithm are contained in figures~\ref{figure 2}-\ref{figure 4}, where in the 
long time limit, averages of the Markov matrix simulations for finite, closed domains are compared to the numerical evaluation of the exact invariant probability distributions in the same potentials, equation~(\ref{eq:boltzmann}).

\section{Mean squared displacement}
We first reproduce the well known result for the ensemble averaged 
mean squared displacement,
which is defined as:
\begin{equation}\label{eq:msd}
	\mathrm{MSD}(t)= \langle (x(t)-x(0))^2 \rangle=\sum_i i^2 \mathrm{P}_i(t)\; ,
\end{equation}
where the latter is correct only for an ensemble of trajectories starting at
$i=0$, i.e., $\mathrm{P}_i(0)=\delta_{i,0}$.
The disorder-averages $\widetilde{\cdot}$ are taken over realisations of 
the potential landscape and in an ensemble average $\langle\cdot\rangle$ over the thermal
noises. In a trajectory-simulation, one would, for
every potential landscape, run a large number of trajectories with their own
thermal noises. In the Markov matrix approach, the ensemble 
average over the thermal
noise is already built in. 

\begin{figure}
\centering
\includegraphics[width=0.48\textwidth]{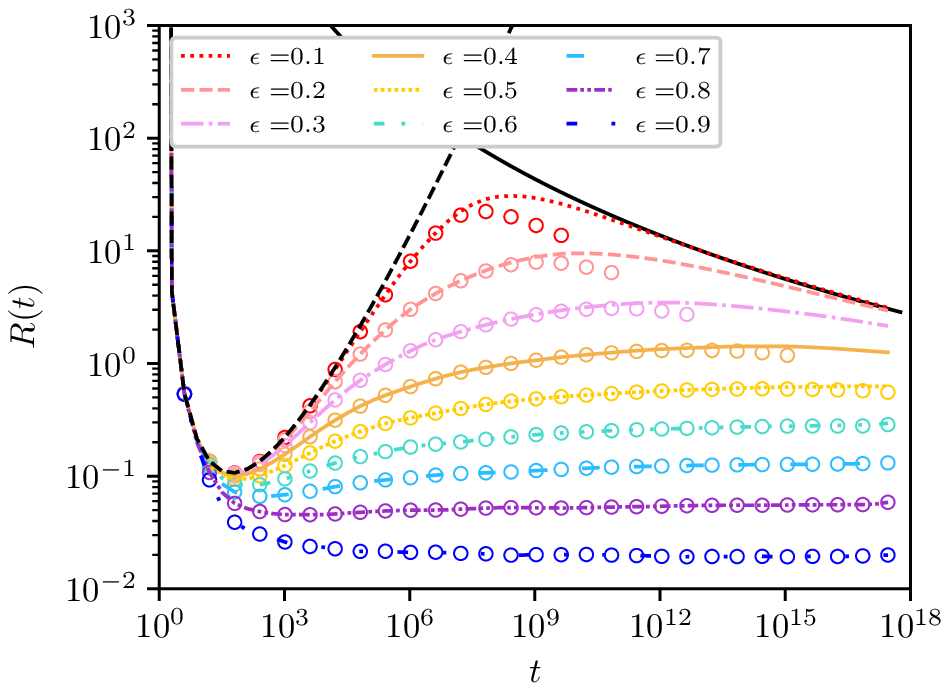}
\includegraphics[width=0.48\textwidth]{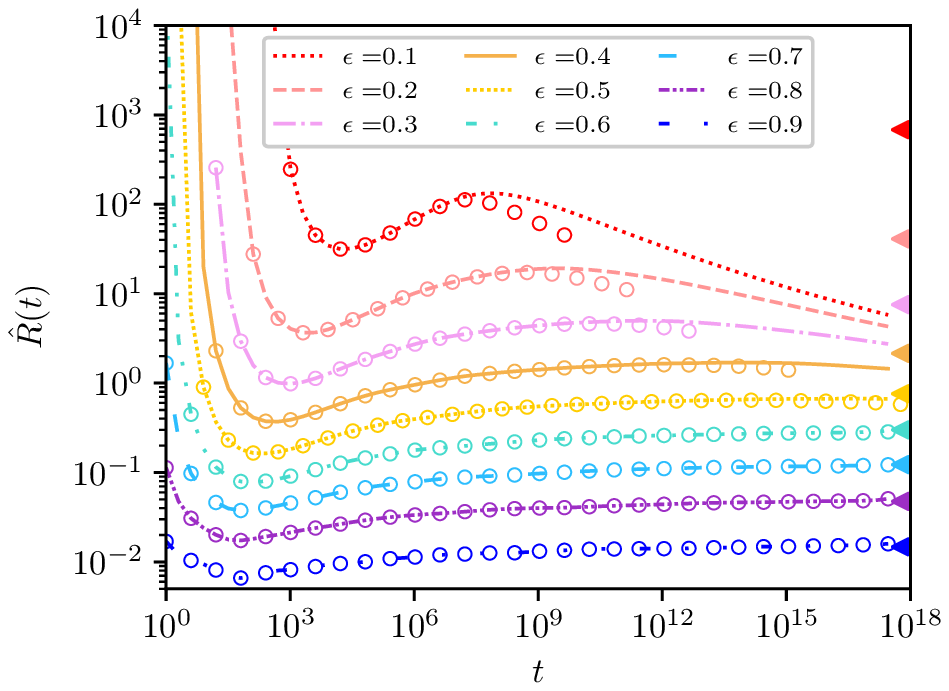}
\caption{\label{figure 2}
Left: disorder-averages of $R(t)=\widetilde{\mathrm{MSD}(t)}/\ln^4 t$, the mean squared displacement divided by $\ln^4 t$ with varying $\epsilon$ as indicated. We compare numerical results obtained by powers of the Markov matrix for open systems 
where (symbols) only the finite range $[-L:L]$ is modelled ($L = 4096$), to results obtained for closed systems (lines). We stop the numerical simulation at times when the computation reaches its limit as explained in the text. The black dashed curve shows how classical diffusion, where the $\mathrm{MSD}(t)\propto t$ looks in this representation, and the solid black line shows a constant MSD, i.e., $1/\ln^4 t$ relation (which is the asymptotic result for closed systems). Right: the same as the left panel but for $\hat{R}(t)=\widetilde{\mathrm{MSD}(t)}/\ln^4(t/\tau)$. The coloured triangles on the right side show the values of the pre-factor in equation~(\ref{msdfunct}).}
\end{figure}

The numerical results shown in figure~\ref{figure 2} are in excellent agreement
with the theoretical prediction $\widetilde{\mathrm{MSD}(t)}\propto \ln^4 t$ \cite{bouchaudetal,Fisher2001}, if time $t$ is
sufficiently large, namely

\begin{equation}\label{msdfunct}
\widetilde{\langle x^{2}(t) \rangle} \simeq \frac{61}{180} \lambda^{2}\ln^{4}\left(\frac{t}{\tau}\right),
\end{equation}
where the length scale $\lambda$ and time scale $\tau$ defined as 

\begin{equation}\label{coeffics}
\lambda=\frac{\eta^{2}D_{\mathrm{eff}}^{2}}{\gamma}, \quad \tau=\frac{\eta^{4}D_{\mathrm{eff}}^{3}}{\gamma^{2}}, \quad D_{\mathrm{eff}}=\left(\frac{180}{61}\right)^{\frac{1}{4}} \frac{k_{B}T}{\eta}.
\end{equation}
Here, $\eta$ is the friction coefficient and $\gamma$ denotes the strength of the disorder. Note that in all numerical simulations we use $\eta=\gamma=k_{B}=1$.
For smaller $t$ we observe some deviations from the 
asymptotic behaviour, which depend on the temperature $T$. This is emphasised
in figure~\ref{figure 2}, where we divide the numerically determined disorder-averaged MSD by
the asymptotic behaviour, $R(t)=\widetilde{\mathrm{MSD}(t)}/\ln^4 t$. 
Not only does the constant of proportionality depend
on $T$ but also the speed of convergence: For 
$\epsilon= 0.6$, i.e., $k_{B}T \approx 0.7$, 
the asymptotic behaviour is reached in the shortest
time. For other values of $k_{B}T$, either the free diffusion (also shown as black line) 
or the deterministic attraction of the deeper wells (finite MSD) dominate the short term behaviour.

This exercise leads us to the conclusion that if we want to observe asymptotic
properties in the shortest simulation time, we should 
use $\epsilon$ in a range of values of 0.6-0.8, or, respectively, 
$k_{B}T\approx0.7-0.5$. However, we will usually perform our numerical simulations for a whole range of $\epsilon$-values. 

\section{More time dependent properties}
\begin{figure}
	\includegraphics[width=0.5\textwidth]{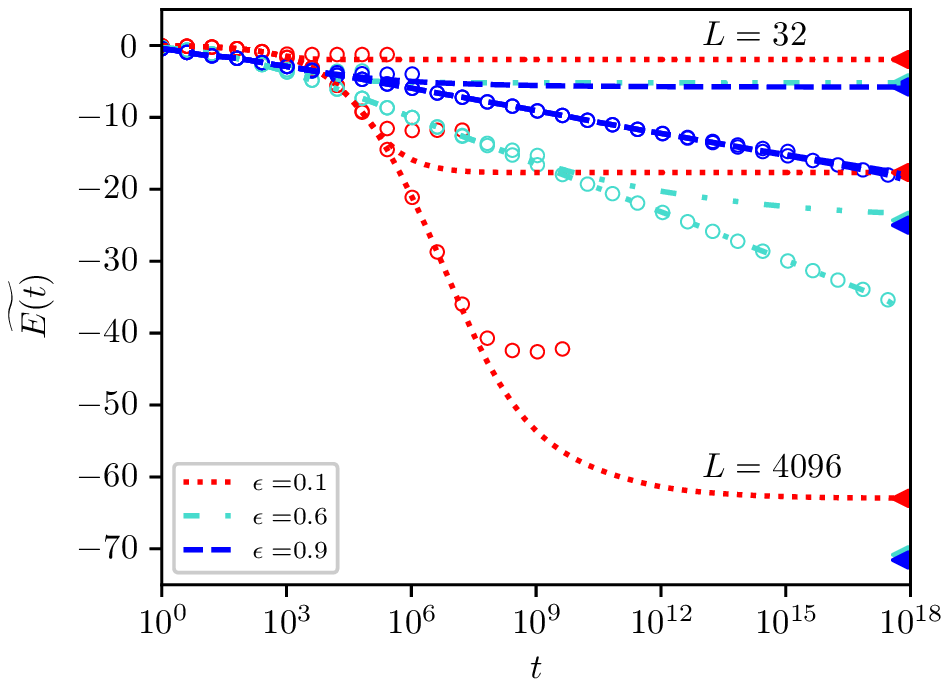}
	\includegraphics[width=0.5\textwidth]{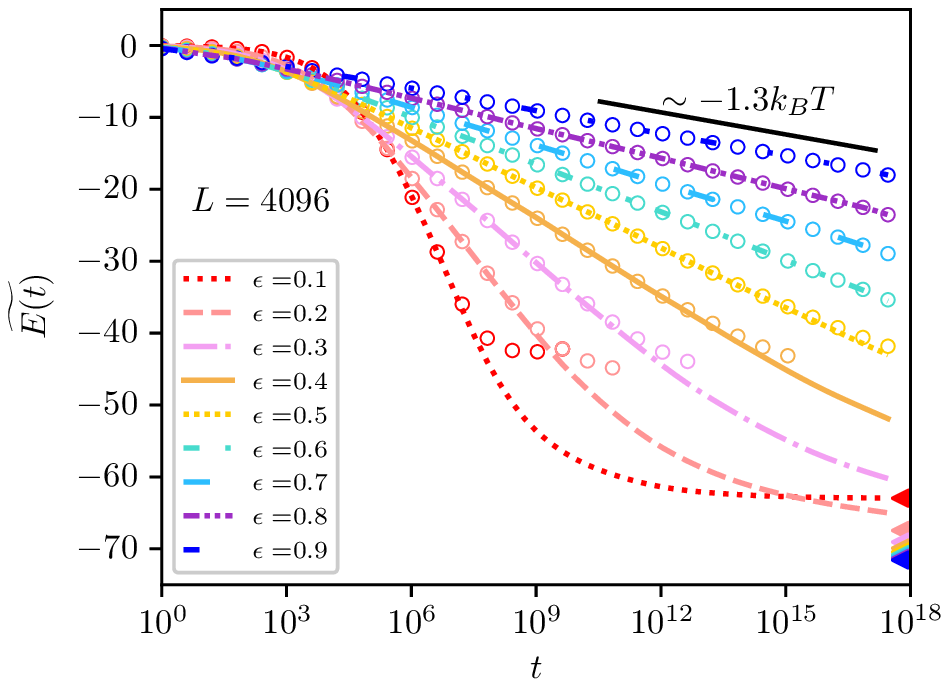}
	\caption{\label{figure 3}
		Time dependent mean potential energy of the Sinai model, obtained from time dependent
		probability distributions through the Markov matrix algorithm, and as
		disorder-average over 10000 realisations of the potential. Left: for different size of the interval $[-L, L]$ ($L = 32, 512, 4096$). Right: for $L = 4096$, with different $\epsilon$ (temperature). In both panels, circles represent the corresponding result for an open system and lines for a reflecting boundary condition on both sides of the interval, and the coloured triangles on the right side show the equilibrium distribution results (Boltzmann distribution). The short black line in the right panel shows a slope  with coefficient $-1.3 k_{B}T$. However, this relation appears to hold only for $T<1$.} 
\end{figure}

\begin{figure}
\includegraphics[width=0.5\textwidth]{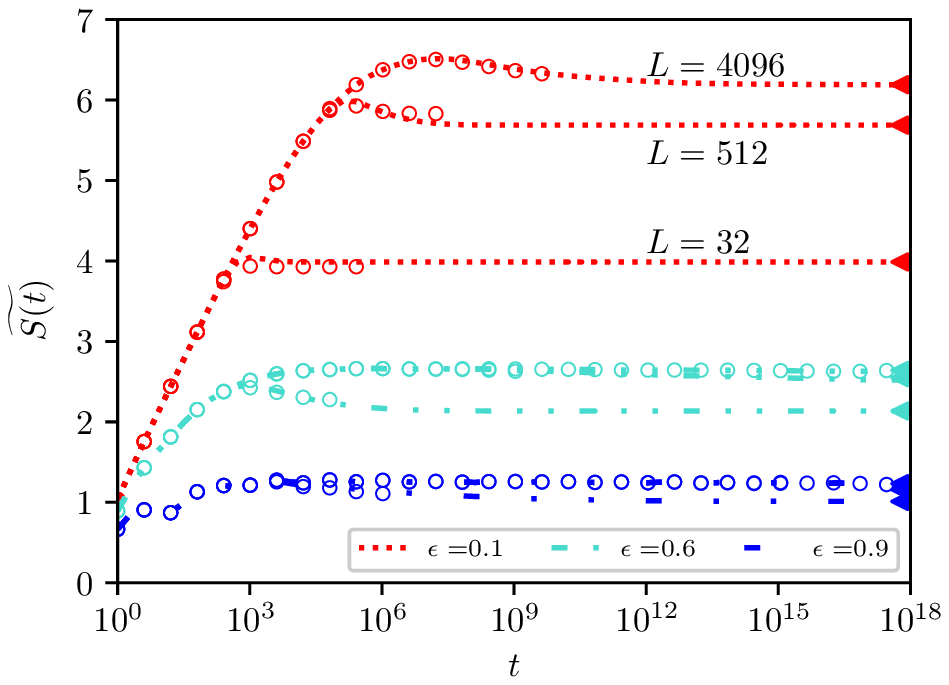}
\includegraphics[width=0.5\textwidth]{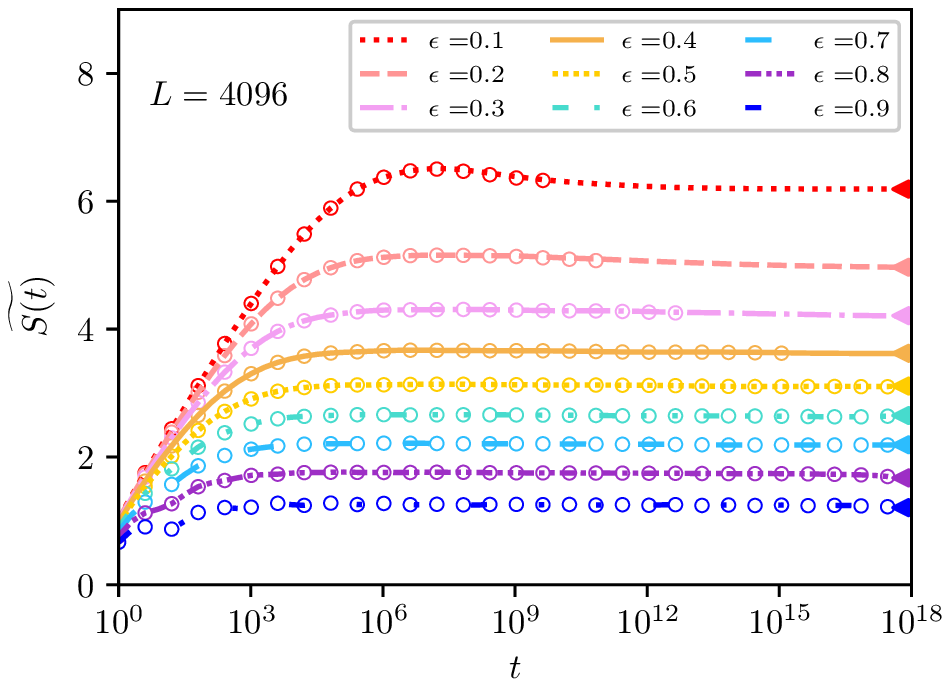}
\caption{\label{figure 4}
Time dependent Shannon entropy of the probability distributions obtained through the Markov matrix algorithm, and as disorder-average over 10000 realisations of the potential, with different $\epsilon$ (temperature). Left: Different lines in the same color code are for different interval lengths ($L = 32, 512, 4096$). Right: for a finite domain with size $L = 4096$ but different temperatures. In both panels, circles represent the corresponding results for an open system and lines for the system with a reflecting boundary condition on both sides, and the coloured triangles on the right end represent the equilibrium values for closed systems (Boltzmann distribution).}
\end{figure}

On an infinite lattice, there is no stationary state. 
Due to the randomness of
the potential in the Sinai model, which has only a statistical 
self-similarity, we cannot expect 
some simple behaviour here. In order to gain insight, 
we calculate the time dependent mean potential energy,
\begin{equation}\label{eq:Energy}
	E(t) = \sum_i V_i \mathrm{P}_i(t),
\end{equation}
and the time dependent Shannon entropy of the probability distribution
$\mathrm{P}_i(t)$,
\begin{equation}\label{eq:Shannon}
	S(t) = - \sum_i \mathrm{P}_i(t) \ln \mathrm{P}_i(t),
\end{equation}
where the sums extend over the whole infinite lattice and $0\ln 0 = 0$.
We study both quantities as  averages over many realisations of the
potential landscape. Numerical results obtained by the Markov Matrix method 
are shown in figures~\ref{figure 3} and \ref{figure 4}.

\begin{figure}
\centering
\includegraphics[width=0.75\textwidth]{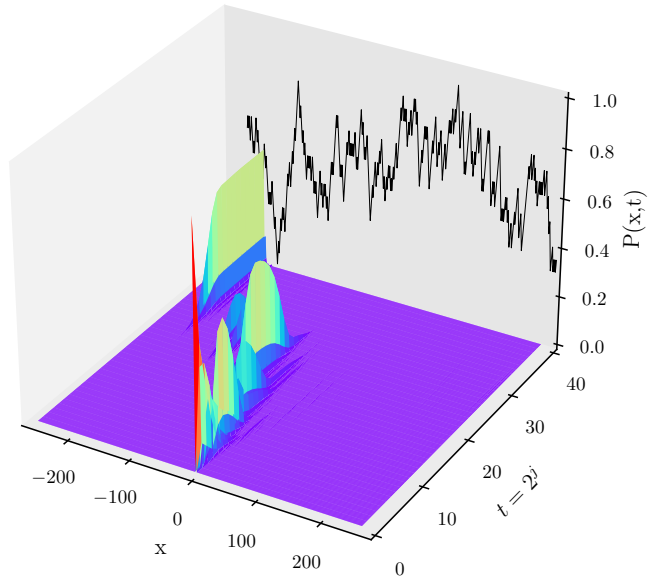}
\caption{\label{figure 5} 
Example of the time dependent probability distribution up to time $t = 2^{40}$ for a single arbitrarily chosen potential ($\epsilon=0.6$). One clearly sees the localisation of probability in
the deepest potential well inside the exploration range, when
the initial distribution is $\mathrm{P}_{0}(0) = 1$ and $\mathrm{P}_{i\neq0}(0) = 0$. The black line represents the random potential and the initial delta-function distribution on $x=0$ is shown in red.}
\end{figure}

The mean potential energy drops as a function of time. This is to be expected,
since, the longer the particle moves through the
potential landscape, it will typically get trapped in even deeper potential wells. This is also exemplified by
a sequence of snapshots of
the time dependent probability distributions $\mathrm{P}_i(t)$ in figure~\ref{figure 5}. In every instance, most of
the probability is concentrated in deep potential wells, but when time goes
on, deeper and deeper wells are explored, and more shallow wells are
vacated. Empirically, we observe the following behaviour:
\begin{equation}
	\widetilde{E(t)} \approx - c(T) \ln t,
\end{equation} 
with a temperature dependent pre-factor $c(T)$. Numerics indicates that this pre-factor is $c(T)\approx 1.3 k_{B}T$ for $T<1$ (see figure~\ref{figure 3}, right panel). This observation can be explained as follows. Consider a particle explores distance $x(t)$. Then, since $V(x)$ is Brownian motion we have $V_{min}(x) \propto \sqrt{x(t)}$  and since $x(t)$ goes like $\ln^2 t$, we get the mentioned scaling of $\widetilde{E(t)}$ with $\ln t$. This means that $\widetilde{E(t)}$ is controlled by the minimum of the potential explored by the particle in time $t$, which is in agreement with other results in this paper. 

Despite this non-stationarity, the snapshots of the time dependent density
show some similarity in their clustering (see figure~\ref{figure 5}). 
This clustering in quantified by the Shannon
entropy, equation~(\ref{eq:Shannon}). 
For a uniform distribution over $N$ lattice points, $\widetilde{S}=\ln N$,
whereas for the initial $\delta$-peak it is $\widetilde{S}=0$. Numerically, we observe 
convergence to an $\epsilon$-dependent constant, $\widetilde{S}(t\to\infty)=const.$. 
Its numerical value suggests that the density asymptotically concentrates on
a few lattice points. Figure \ref{figure 5} makes it plausible that this clustering of
the probability density takes place inside 
the deepest potential well in the explored region of the
given potential.

So the intuition which we gain from this low-temperature 
non-equilibrium simulation is that 
at time $t$ the particles explore a range $L_{\mathrm{eff}}$ 
of the potential which scales like
$\ln^2 t$, and that they settle down in the absolute minimum of this part of 
the potential. As time goes on, the range grows, and therefore new and even
deeper absolute minima are explored. This leads to a decrease of the mean
potential energy but to constancy of the entropy. The energy barrier after release from $x=0$ at $t=0$ is of order $\sqrt{\gamma x}$, and the time required to cross this barrier is given by the Arrhenius law, $t\simeq\tau\mathrm{exp}\left(\sqrt{\gamma x}/k_{B}T\right)$. Here $\gamma$ represents the strength of the disorder and $\tau$ defines a fundamental time scale. According to these scaling relations, after
the time $t$ the particle typically has covered the distance $\langle x^{2}(t)\rangle\simeq\ln^{4}(t/\tau)$, see \cite{bouchaudetal}.

\subsection{Golosov effect}

We now address the localisation effect of the Sinai model described by Golosov in \cite{Golosov} (see also \cite{bouchaudetal}) by computing the standard deviation $\sigma(t)=\sqrt{\langle x^{2}(t)\rangle-\langle x(t)\rangle^2}$, in a semi-infinite domain where a particle starts its motion from $x=0$ with a reflecting boundary condition at the origin. In \cite{Golosov} it was stated that this $\sigma(t)$ would asymptotically for large $t$ approach a finite value (disorder dependent), while the mean value $\langle x(t)\rangle$ is governed by a function $m(t)$ which tracks the deepest well 
in the explored range of the random potential. More precisely, he proved that the disorder-averaged relative distance $x(t)-m(t)$, in the long time limit converges towards a limit distribution. Hence, 
in the Golosov scenario the width of a packet of particles within a single realisation of the random potential does not grow with time, in the long time limit. However, taking an average over different disorders leads to a divergent standard deviation. 
This is due to those configurations of the random potential in which there are more than a single deep well, and where the particles usually are localised in different spatially separated minima, which leads $\sigma(t)$ to diverge in the infinite-time limit (see also \cite{Chave1999,Radons2004}). Indeed, it has been shown that the disorder-averaged standard deviation of the Sinai model has the following long-time asymptotic \cite{Laloux1998, Monthus2002, Doussal1999},

\begin{equation}
\label{eq:standdiv}
\widetilde{\sigma(t)}\simeq\frac{61}{180}\lambda\ln^{3/2}\left(\frac{t}{\tau}\right),
\end{equation}
where $\lambda$ and $\gamma$ defined as equation~(\ref{coeffics}).
\begin{figure}
\centering
\includegraphics[width=0.49\textwidth]{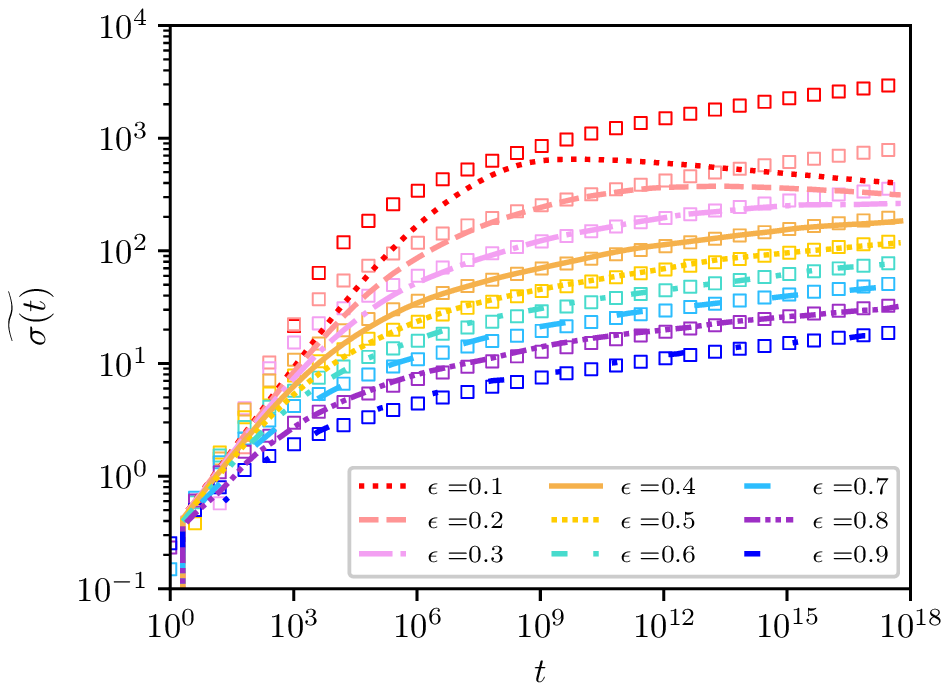}
\includegraphics[width=0.49\textwidth]{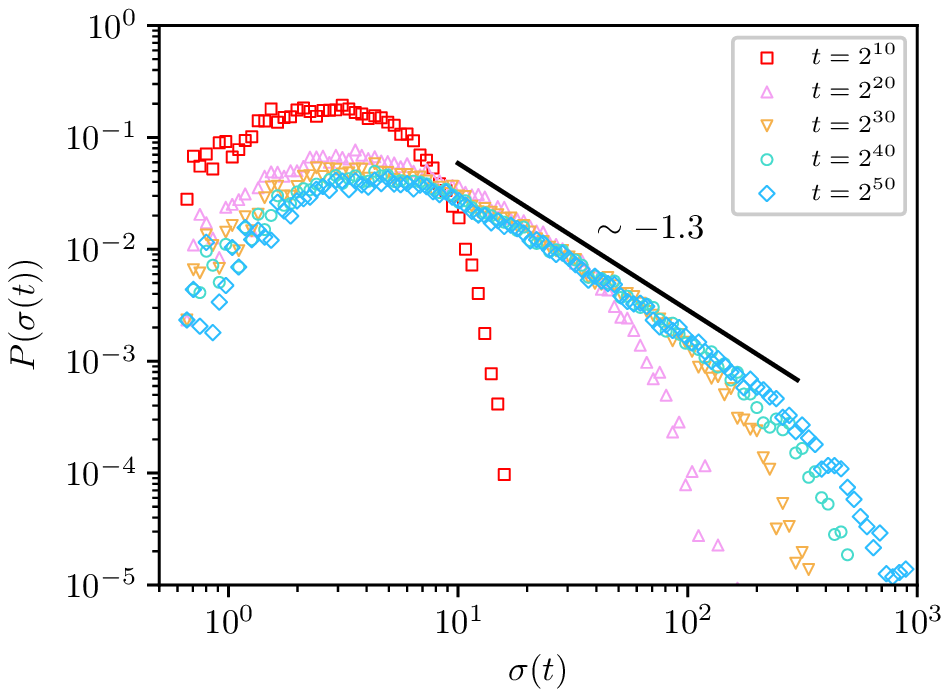}
\caption{\label{figure 6} Left: disorder-averaged standard deviation $\widetilde{\sigma(t)}$ for different values of $\epsilon$ of Sinai model in a semi-infinite domain with a reflecting boundary condition at the origin. Lines represent simulation results obtained by the Markov matrix approach and square symbols show the analytical asymptotic behaviour~(\ref{eq:standdiv}). Right: PDF of the standard deviation
for $\epsilon = 0.6$ at different times in a semi-infinite domain obtained from $10000$ realisation of the random potential. As a guide to the eye the short black line shows a slope  of $-1.3$.
Here we use $L=2^{13}$. }
\end{figure}
In figure~\ref{figure 6} (left panel) we show the results of disorder-averaged standard deviation for different temperatures (see equation~(\ref{eq:temp})) and compare it with the asymptotic behaviour (\ref{eq:standdiv}).
As can be seen, for large $\epsilon$ (low $T$), 
there is a very good agreement between the analytical prediction and the numerical data. However, for
small $\epsilon$ (high $T$), since this is affected by the finiteness
of system size $L$, we observe a deviation from the theory.

Moreover, in the right panel of figure~\ref{figure 6} we demonstrate the PDF of the standard deviation for $\epsilon=0.6$ at different times. As can be seen, at long times the PDF of $\sigma(t)$
is almost time independent, which is in agreement with Golosov's
theorem, with a power-law decay whose power guarantees normalisation but, without cut-off, would yield a diverging mean value. The cut-off at large $\sigma(t)$, however, is a function of $t$ since $\sigma(t)$ in every individual potential is strictly bounded by the largest distance 
a particle can travel in time $t$ from the origin, and this propagation is extremely slow. Therefore, the mean values of these PDFs at any finite time are finite, but slowly growing in time and eventually diverging, in full agreement with  equation~(\ref{eq:standdiv}). In this sense, the statement of Golosov's theorem about an asymptotic shape of the distribution of $\sigma(t)$ and equation~(\ref{eq:standdiv}) are not in contradiction, as it is illustrated by our numerics, due to the power-law tail with time dependent cut-off.

 Our analysis in terms of entropy, however, shows that entropy $\widetilde{S(t)}$ converges to a finite value, which  characterises the localisation of the thermal particles regardless of whether this takes place in a single or in multiple wells, and which is also insensitive to the distance between these wells. Therefore, the entropy is a much more suitable indicator for localisation than the standard deviation.

\section{Equivalence of equilibrium and non-equilibrium	dynamics\label{sec:equivalence}} 

In this section we want to prove our claim that the unbounded motion of the infinite 
system is always in close vicinity of an equilibrium solution of a system whose 
size is given by the average exploration range of the unbounded motion at the respective time.

For the open systems, we therefore fix the range $[-L,L]$ on which we model its time evolution, and 
we stop the iterations when probability starts to leak out through the open boundaries. 
We then calculate the entropy and the energy of this probability distribution in the given potential, and again perform an average over 10000 realisations of the random potential.
These values will then be compared to those calculated with the Boltzmann distribution equation~(\ref{eq:boltzmann}) in the same potentials. 

\begin{figure}
\centering
\includegraphics[width=0.6\textwidth]{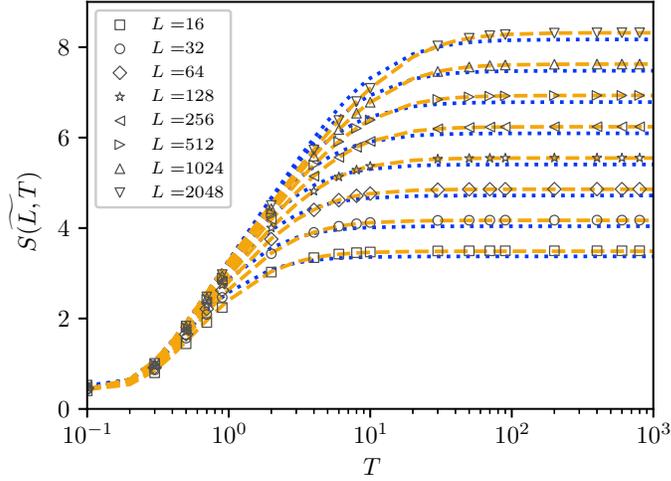}
\caption{\label{figure 7}
Disorder-average of the Shannon entropy of the equilibrium distributions 
of finite $L$ Sinai models. The maximum values are $\widetilde{S}=\ln(2L)$ and represent the
uniform high-temperature limit. The saturation for low temperatures reflects
the localisation in the deepest well. Symbols represent the corresponding results to the Boltzmann distribution, dashed and dotted lines show the results for reflecting boundary condition and open system, respectively. The maximum values of the entropy for the open system are $\widetilde{S}=\ln(2L_{\mathrm{eff}})$, in which $L_{\mathrm{eff}}\approx0.87L$.}
\end{figure}

\begin{figure}
\includegraphics[width=0.5\textwidth]{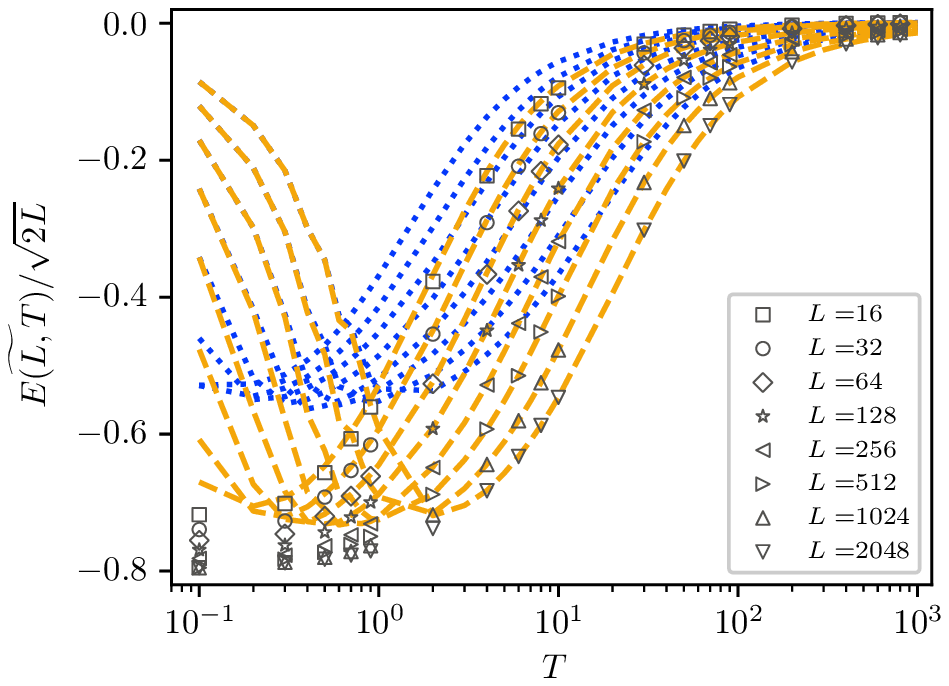}
\includegraphics[width=0.5\textwidth]{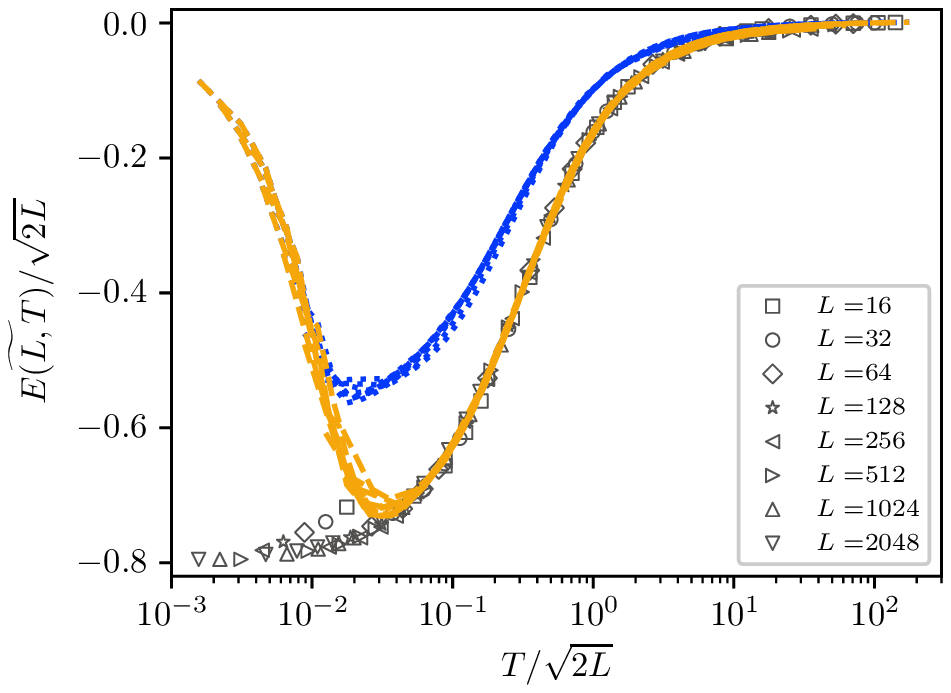}
\caption{\label{figure 8}
Disorder-average of the mean potential energy in equilibrium of Sinai
models of different size $L$. We rescale the energy with $1/\sqrt{2L}$ since
this is how the well depths scale. Symbols represent the corresponding results to the Boltzmann distribution, blue dotted and orange dashed lines show the simulation results for an open system and a reflecting boundary condition, respectively. As can be seen, at low temperatures in the system with reflecting boundaries, the mean potential energy deviates from the values of the equilibrium due to the numerical error of the Markov matrix algorithm. The same phenomenon is observed for the open system.}
\end{figure}

The dependence of both entropy and mean potential energy on the
temperature for different lattice sizes $L$ are shown in 
figures~\ref{figure 7} and 
\ref{figure 8}, respectively.

Following the above argument, the random potential is given as a random walk, whose deviation from
the origin scales like $\sqrt{L}$. Assuming the same scaling for the deepest
potential well on the finite range $2L$, we expect the equilibrium mean
potential energy, which for small $T$ is dominated by exactly the deepest
well, to scale like $-\sqrt{2L}$. This is similar to the above argument, in which instead of $t$ we have $L$.  Indeed, when re-scaling $\widetilde{E}$ in
this way, we obtain a nice data collapse for very small and very large temperatures, see
figure \ref{figure 8}. Actually, the data collapse for $T\gg
1$ is no surprise, since the mean potential energy is zero in this limit of a
uniform distribution, independent of $L$. For the intermediate range, one
finds a data collapse as well when re-scaling also temperature by $1/\sqrt{2L}$, see figure~\ref{figure 8}. 

This result can be explained as follows. Consider a system with size $[-L, L]$. We order the potentials at each lattice point, from minimum to maximum. The two lowest potential traps are called $V_{min}$ and $V_{next}$. We can add more minima to the argument below, but we use only two deepest valleys and assume a Boltzmann distribution. In this approximation, 
the mean energy reads
\begin{equation}
    \langle E\rangle = \frac{1}{Z}\left(V_{min}\mathrm{e}^{- V_{min}/k_{B}T} + V_{next} \mathrm{e}^{- V_{next}/k_{B}T}\right),
\end{equation}
where $Z$ is the partition function of this two level system,
$Z= \mathrm{exp}( - V_{min}/k_{B}T) + \mathrm{exp}( - V_{next}/k_{B}T)$. Note that $V_{min}$ and $V_{next}$ are typically negative. Now, the usual argument is that
$V_{min}= c_{min} \sqrt{L}$ and $V_{next} = c_{next} \sqrt{L}$, in which
$c_{min}$ and $c_{next}$ are statistically independent of $L$, and specific to the system.
Inserting this in the ensemble average gives us
\begin{equation}
    \langle E\rangle = \frac{\sqrt{L}}{Z}\left(\mathrm{e}^{-c_{min}\frac{\sqrt{L}}{k_{B}T}}
+ \mathrm{e}^{-c_{next}\frac{\sqrt{L}}{k_{B}T}}\right).
\end{equation}
Therefore, plotting of $\langle E\rangle/\sqrt{L}$ versus $T/\sqrt{L}$ is $L$ independent, as is shown in figure~\ref{figure 8} (right panel).
As mentioned, we can also add other minima, and the same trick will work.
Furthermore, the ensemble average over the disorder when added will give a non-random result (the result will not depend
on the specific values of $c_{min}$ and $c_{next}$).
 
The entropy, in contrast, turns out to be independent of system size $L$ in
the $T\to 0$ limit, since only the shape of the deepest well and its degeneracy matters.
The shape can be assumed to be independent of $L$. There is a very slow increase with $L$ since the
bigger $L$ the larger is the probability that there is a second, independent
well with the same depth and hence a larger degeneracy of the minimum.

As said, the open system results shown in figures~\ref{figure 7} and \ref{figure 8} show mean values taken at the time when the total probability in the range $[-L,L]$ starts to drop, which means that the support of the time dependent probabilities starts to extend beyond the range $[-L,L]$. Thus the time dependent exploration range $L_{\mathrm{eff}}$ becomes larger than the modelled domain, and this occurs after a time given by $L_{\mathrm{eff}}\propto\ln^2 t$. Hence, for the open system, the behaviour $\widetilde{E(t)} \propto -\ln t$ observed above translates into $\widetilde{E(t)}\propto \sqrt{L_{\mathrm{eff}}(t)}$ in which $L_{\mathrm{eff}}\approx0.57L$, thus is consistent with the equilibrium result $\widetilde{E(L)}\propto\sqrt{L}$.

This analysis therefore suggests that the non-equilibrium dynamics relaxes close to equilibrium in the
finite range it has explored so far. This is an evident consequence of the
time scale separation of the problem: Ultraslow diffusion across large scales,
but thermal noises which are strong enough to have fast local relaxation.

\section{Three temperature regimes and scaling}

For two types of dynamics, finite time simulation for finite $L$ system
and an open system, it is reasonable to distinguish the low, medium,
and high temperature regimes separately. In the low temperature limit, both
types of dynamics are related through the relationship of system size $L$ and
the exploration horizon $L_{\mathrm{eff}}$, as explained above. 

In the high temperature limit, the equilibrium distribution of a finite size
system is uniform, hence the mean potential energy is zero, and the entropy
of this distribution is $\ln L$. In the non-equilibrium situation, high
temperature implies irrelevance of the potential and hence free diffusion with
its well known properties. In particular, while the mean potential energy 
remains 0 for all times, the entropy increases as $\ln t$ without upper
bound\footnote{The Shannon entropy of a Gaussian is proportional to the
logarithm of its standard deviation.}. 
Employing again the scaling of the
MSD in time in order to translate time into system size, we now have
$L_{\mathrm{eff}}\propto \sqrt{t}$. The finite-time entropy of the infinite
system therefore depends on the effective system size as: $\widetilde{S}\simeq \ln
L_{\mathrm{eff}}^2 \approx 2 \ln L_{\mathrm{eff}}$. Apart from the mismatch of the
prefactor, the difference of the distributions (Gaussian versus uniform) make
it evident that  in the high temperature regime, there
is no equivalence of the non-equilibrium and equilibrium
behaviour. 

Finally, let us shortly discuss the regime of intermediate energies.
We call energies intermediate, if the equilibrium density is concentrated in
many more wells of the potential than the deepest one but is still localised. 
Due to the scaling of the energy landscape in system size, we expect that
equilibrium systems of different size are equivalent when the temperature
divided by $\sqrt{L}$ is the same. 
Actually, this is what we observe in
figure~\ref{figure 8} 
in the central range of the temperature scale. 
Note that whenever we
compare energies for different system sizes, 
we normalise them by $\sqrt{L}$ since this is the expectation
value for the deepest well.

\section{Relation to infinite densities}

\begin{figure}
\centering
\includegraphics[width=0.6\textwidth]{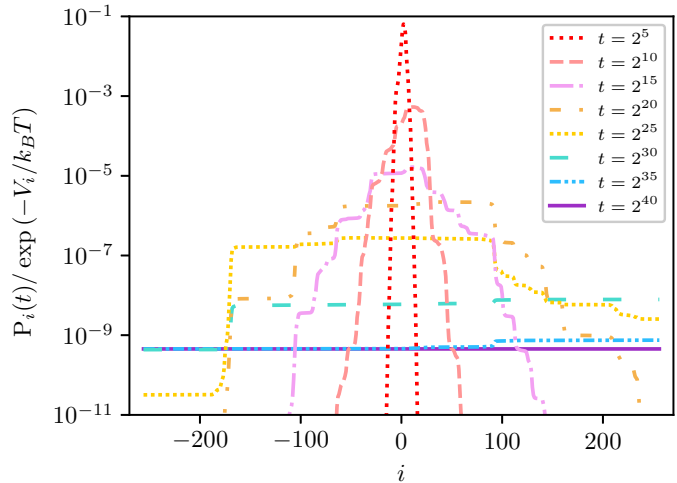}
\caption{\label{figure 9} 
Ratios of the empirical probabilities $r_i=\mathrm{P}_{i}(t)/e^{-V_{i}/k_{B}T}$ after
different times $t$. Evidently, for every time there is a core region where $r_i = O(1)$,
and exterior regions where $r_i\approx 0$, with very narrow transitions (except at short times since there was not sufficient time to relax to a flat shape). 
Different lines represent different times $t=2^j$ up to $j=40$. Here we use $\epsilon=0.6$ (or $k_{B}T\approx0.7$) and $L=256$.}
\end{figure}

Recently, there has been much progress in the study of systems with
asymptotically flat potentials. The Boltzmann distribution $e^{-V(x)/k_{B}T}$ for
flat potentials is not normalisable, since it is constant for large $|x|$. 
This gives rise to the notion of infinite densities and to approaches to 
deal with these, see \cite{Aghion2019, Aghion2020, Dechant2011, Farago2021, Siler2018, Defaveri2020, Anteneodo2021, Radons1996, Akimoto2010, Akimoto2013, Akimoto2020, Kantz2017, Kantz2021}. Inspired by these 
works, we study here the ratio $r_i$ of the time 
dependent, numerically generated
probability $\mathrm{P}_i(t)$ and of the non-normalised Boltzmann distribution 
on the lattice, $\mathrm{P}_{i}^{B}=e^{-V_i/k_{B}T}$, namely $r_i=\mathrm{P}_{i}(t)/\mathrm{P}_{i}^{B}$. For large $|i|$, $\mathrm{P}_i(t)=0$, and so is
this ratio $r_i$. However, inside the exploration range, this 
ratio is an $i$-independent constant $c(t)$ 
with a very narrow range. We eliminate this constant by summing up
the product $\mathrm{P}_i(t)e^{V_i/k_{B}T}$ which is essentially (but not precisely) 
the norm of the Boltzmann
distribution inside the exploration range. Hence we use this value 
as normalisation of the Boltzmann distribution,
see figure~\ref{figure 9} for an illustration for a single random potential. 
This is another way to verify the validity of the
concept of local equilibrium: Ensemble of the open
system on the infinite lattice explores some range which its inside almost
always, to some good approximation, represents 
the equilibrium distribution of a finite, closed system, 
while outside the probability to find the
particle is essentially zero, with a narrow transition in between. Note that
time steps between the different snapshots in figure~\ref{figure 9} are
exponentials, $t=2^j$. 

\begin{figure}
\includegraphics[width=0.5\textwidth]{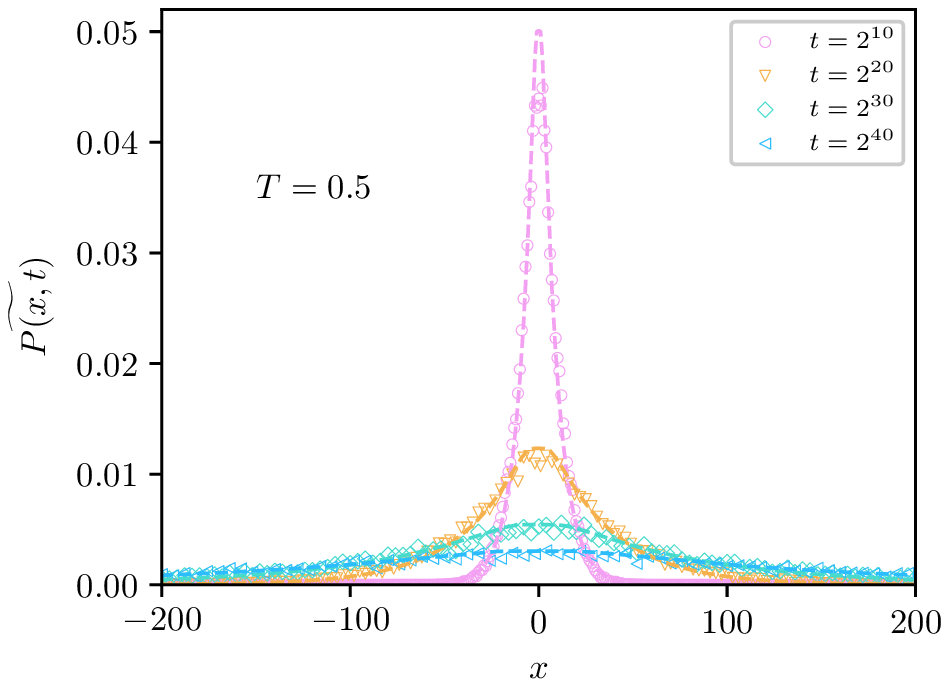}
\includegraphics[width=0.5\textwidth]{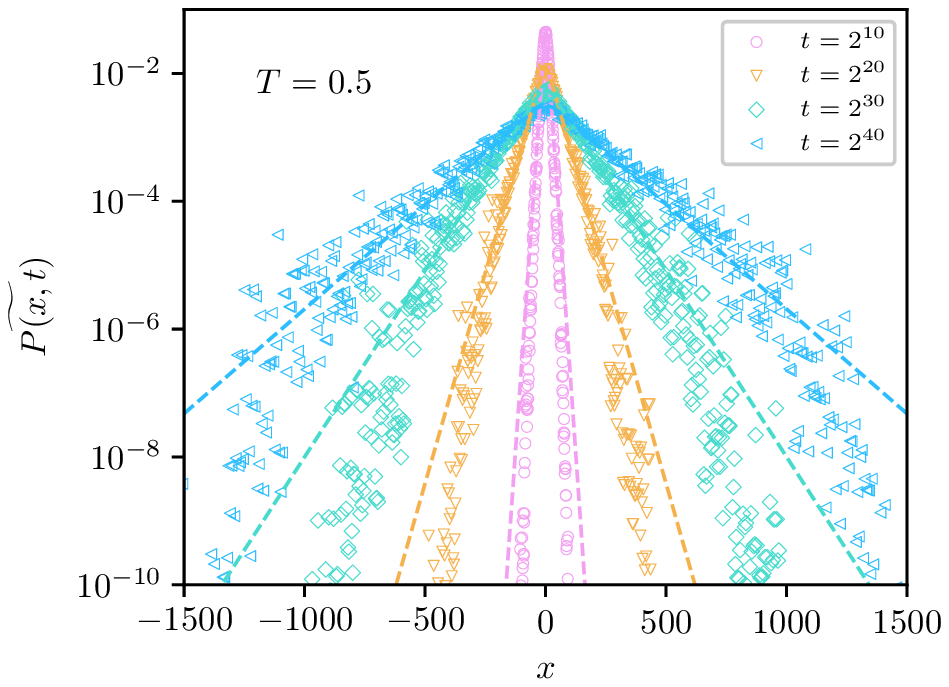}\\
\includegraphics[width=0.5\textwidth]{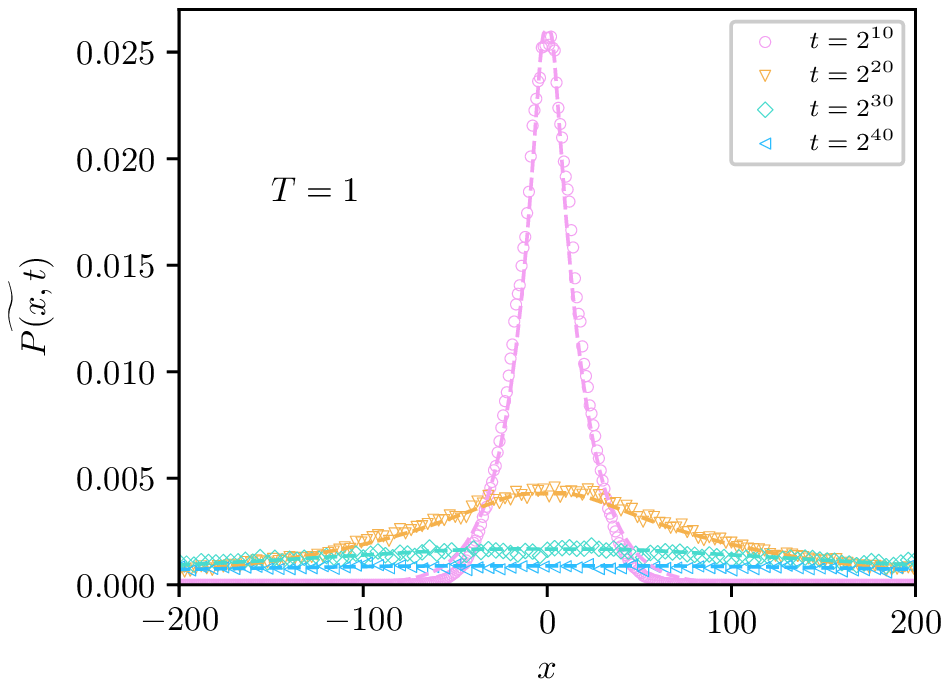}
\includegraphics[width=0.5\textwidth]{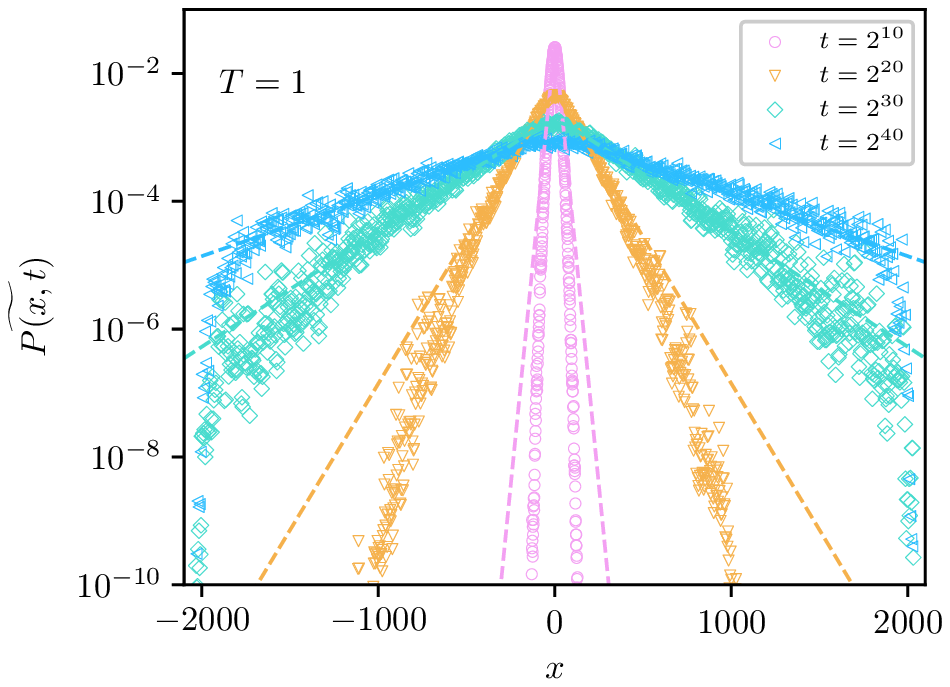}
\caption{\label{figure 10} 
PDFs of the Sinai diffusion at different times. Symbols represent the simulation results obtained through the Markov matrix algorithm for an open system $[-L,L]$ with length $L=2048$, and dashed lines 
correspond to the analytical solution (\ref{eq:sinaipdf}).}
\end{figure}

We finally study the time dependent probability distributions $\mathrm{P}_i(t)$ averaged over random potentials. For continuous space and time, 
the long time behaviour of the Sinai diffusion PDF is as follows \cite{Comtet1998, Kesten1986, Nauenberg1985, Bunde1988} 

\begin{equation}\label{eq:sinaipdf}
\widetilde{\mathrm{P}(x,t)} \sim \frac{4}{\pi \lambda \ln^{2}(t/\tau)} \sum_{n=0}^{\infty} \frac{(-1)^n}{2n+1}\exp\left(-\frac{(2n+1)^2 \pi^2|x|}{4\lambda \ln^{2}(t/\tau)}\right),  
\end{equation}
where $\lambda$  and $\tau$ defined as equation~(\ref{coeffics}).
We compare this analytical results to our numerics for different temperatures $T$ in figure~\ref{figure 10}. On logarithmic scales, for $T = 0.5$ we see some mismatch at tails but an overall good agreement\footnote{Note that, in order to compare the analytical solution with our numerics, we set $\tau\to8\tau$ in equation~(\ref{eq:sinaipdf}). The discrepancy might be due to different definitions of the parameters such as $D_{\rm eff}$ in the lattice hopping model.}.

\section{Conclusions}

Sinai diffusion belongs to the class of classical hard problems in statistical mechanics
and numerous questions on its detailed behaviour are still open. 
Here we consider the local equilibrium behaviour of a thermal particle in presence of a random force field. We make use of a very fast numerical scheme to study the time dependent densities, mean potential energy and the Shannon entropy as well as the mean squared displacement of the ensemble particles in a bounded domain with reflecting boundaries and in an open system by averaging over $10^4$ random potentials with a Brownian path. 
Our analysis in terms of the time dependent densities shows that while the system is in a non-equilibrium state, as manifested in the time dependent mean square displacement, still it exhibits
some properties which are inherently related to thermal equilibrium. For example with this insight we could use simple scaling arguments, based on the extreme of the minima to find the energy of the system, see figure~\ref{figure 8}.

In contrast to the setup analysed by Golosov, who considered the
localisation of a particle packet released in the  deepest well of a specific realisation of the random potential we here consider the case when
ensembles of particles are seeded in an arbitrary position of the random
potential and allowed to evolve independently. While Golosov stated that the variance of an ensemble of particles should be approximately constant over time, we find instead that the variance continues to increase even in the long time limit, but that instead the Shannon entropy converges to a constant. Together this implies that the time dependent PDF is localised in a small number of potential wells, but that the distance between wells which are populated at a time usually grows in time.
In agreement with Golosov, we find that
in an ensemble average over different random potentials,
there is convergence in time to a unique distribution with power law
tails but a time dependent cut-off, which hence yields a finite
variance at any finite time.
Comparing the unbounded
motion of the particles in an infinite 1-dimensional domain with the motion in finite, bounded domains with reflecting boundaries we demonstrate
that the unbounded motion is close to the equilibrium state  of a finite
system of growing size at all times. This observation is due to the distinct time scale separation, according to which inside the wells of the
random potential, there is a relatively fast equilibration, while the motion across major potential barriers is ultraslow.  Our results shed new
light on the equlibration behaviour of particle packets in quenched, disordered potential landscapes.

Studying the time averaged spreading characteristics of a non-normalised state would be interesting extensions of the present work. The quantitative characterisation in terms of trajectories, such as time averaged MSD
and width of the particle packet will be of use in the analysis of dynamic
phenomena in strongly disordered energy landscapes. 
For instance, the strong ageing
observed in simulations of single proteins may indicate that protein dynamics may belong to this class of problems \cite{HuX2016}.

Another open problem is how to relate between the infinite densities, namely the fact that within a range the system is in a  non-normalisable Boltzmann-Gibbs state, and the well known disorder average propagation $\widetilde{\mathrm{P}(x,t)}$, equation~(\ref{eq:sinaipdf}).

\section*{Acknowledgements}

AC acknowledges support of the Polish National Agency for Academic Exchange
(NAWA). The support of Israel Science Foundation's grant 1614/21 is acknowledged (EB). RM acknowledges the German Science Foundation (DFG, grant no. ME 1535/12-1) and the Foundation for Polish Science (Fundacja na rzecz Nauki Polskiej, Humboldt Polish Honorary Research Scholarship) for support.

\end{document}